\documentclass[prd,preprint,amsmath,amssymb,nofootinbib]{revtex4-1}
\usepackage[utf8]{inputenc}
\usepackage[T1]{fontenc}
\usepackage[dvipsnames]{xcolor}
\usepackage{mathrsfs}

\usepackage{hyperref}

\newcommand{\ULB}{Universit\'e Libre de Bruxelles and International Solvay Institutes, \\ Campus Plaine, C.P.~231, B-1050 Bruxelles, Belgium}

\newcommand{\UVA}{Department of Physics, University of Virginia, P.O.~Box 400714, Charlottesville, VA 22904-4714, USA}

\begin{document}

\title{Classical and Quantized General-Relativistic Angular Momentum}

\author{Geoffrey Comp\`ere}
\email{gcompere@ulb.ac.be}
\thanks{(Corresponding Author)}
\affiliation{\ULB}

\author{David A.\ Nichols}
\email{david.nichols@virginia.edu}
\affiliation{\UVA}

\date{\today\\ \vspace{1cm}}

\begin{abstract}

\vspace{\baselineskip}
Angular momentum at null infinity has a supertranslation ambiguity from the lack of a preferred Poincar\'e group and a similar ambiguity when the center-of-mass position changes as linear momentum is radiated.
Recently, we noted there is an additional one-parameter ambiguity in the possible definitions of angular momentum and center-of-mass charge. We argue that this one-parameter ambiguity can be resolved by considering the generalized BMS charges that are constructed from local 2-sphere-covariant tensors near null infinity; these supertranslation-covariant charges differ from several expressions currently used. 
Quantizing angular momentum requires a supertranslation-invariant angular momentum in the center-of-mass frame.
We propose one such definition of angular momentum involving nonlocal quantities on the 2-sphere, which could be used to define a quantum notion of general-relativistic angular momentum. 

\vspace{\baselineskip}

Essay written for the Gravity Research Foundation 2021 Awards for Essays on Gravitation.

\end{abstract}

\maketitle

A new twist entered the long history of defining the total angular momentum of asymptotically flat spacetimes 
in general relativity \cite{Bondi:1962px,*Sachs:1962wk,*PhysRev.128.2851,1966PhRv..150.1039T,*1982JMP....23.2410A,*Penrose:1982wp,1975ctf..book.....L,Thorne:1980ru,Ashtekar:1981bq,Dray:1984rfa,Wald:1999wa}. The ``supertranslation ambiguity'' of angular momentum (the lack of a preferred Poincar\'e subgroup of the Bondi-Metzner-Sachs (BMS) group~\cite{Bondi:1962px,*Sachs:1962wk,*PhysRev.128.2851}) and the ``center-of-mass (CM) ambiguity'' (the fact that the CM rest frame will change when linear momentum is radiated) described, e.g., in~\cite{Ashtekar:2019rpv} are well known. Recently, we showed that there is an independent one-parameter family of nonequivalent local definitions of angular momentum and center-of-mass charges in general relativity~\cite{Compere:2019gft,Elhashash:2021iev}. 
The charges in this family all satisfy the following desirable criteria: they vanish in flat spacetime, reduce to the Kerr angular momentum for a rotating black hole in its CM rest frame, are locally constructed from tensors on the celestial sphere in a neighborhood of a cut of null infinity, obey the standard BMS charge algebra at spatial infinity~\cite{Troessaert:2017jcm,*Henneaux:2018cst,Compere:2020lrt}, satisfy BMS flux balance laws at null infinity~\cite{Barnich:2011ty,*Barnich:2011mi,*Distler:2018rwu,Compere:2019gft}, and are consistent with the subleading soft-graviton theorem \cite{Cachazo:2014fwa,*Kapec:2014opa,*Campiglia:2014yka}. 
Moreover, different definitions are used (to date) in different sub-fields of gravity, although they are nonequivalent: for example, for black-hole mergers (like those measured by LIGO and Virgo), the difference can be a $\sim0.1$\% effect~\cite{Elhashash:2021iev}. 

We did not find a compelling reason to prefer one definition over another in~\cite{Compere:2019gft,Elhashash:2021iev} based on the criteria above. 
We propose in this essay that the one-parameter ambiguity is \emph{uniquely} fixed for asymptotically non-radiating spacetimes once we require the angular momentum be part of the larger algebra of generalized BMS charges at spatial infinity as found in \cite{Compere:2020lrt} (which completes the earlier work of~\cite{Barnich:2011mi,Barnich:2011ty,Campiglia:2015yka,Hawking:2016sgy,Compere:2018ylh,Distler:2018rwu,Campiglia:2020qvc}).
The resulting angular momentum differs from two commonly used definitions in~\cite{Thorne:1980ru,Ashtekar:1981bq,Dray:1984rfa,Wald:1999wa} and~\cite{1975ctf..book.....L,Bonga:2018gzr,*Damour:2020tta} (and from the recently proposed definition \cite{Chen:2021szm}), but it agrees with a definition~\cite{Pasterski:2015tva,Hawking:2016sgy} in a trivial CM and supertranslation frame. 
In quantum gravity, the generalized BMS conserved charges at spatial infinity that faithfully represent generalized BMS symmetries~\cite{Compere:2020lrt} will be promoted to quantum operators. The quantized angular momentum can be defined at spatial infinity in a fixed frame, but to define it on null infinity, a new supertranslation- and CM-invariant definition is needed. 
We propose such a definition that is compatible with generalized BMS symmetry, which is necessarily non-local.

We first introduce our notation and conventions.
Asymptotically flat solutions can be given in Bondi~\cite{Bondi:1962px,Sachs:1962wk} gauge $x^\mu=(u,r,x^A)$, where $x^A=(\theta,\phi)$. 
Assuming $g_{uu}=O(r^0)$, the diffeomorphism group is reduced to the group of generalized BMS asymptotic symmetries which consist of supertranslations and super-Lorentz transformations \cite{Barnich:2010eb,Barnich:2011mi,Barnich:2011ty,Campiglia:2015yka,Hawking:2016sgy,Compere:2018ylh,Campiglia:2020qvc}. 
We denote the metric on the celestial sphere by $\gamma_{AB}$ ($D_A$ is its covariant derivative), the shear tensor by $C_{AB}(u,x^C)$ and the Bondi mass aspect by $m(u,x^C)$ [it appears in $g_{uu}=-c^2+2Gm/(c^2 r)+O(r^{-2})$]. 
The Bondi angular-momentum aspect $N_A(u,x^C)$ appears in the metric component 
\begin{equation}
g_{uA} = -\frac{1}{2}D^BC_{AB}+\frac{1}{r}\left( \frac{4G}{3c^2} N_A - \frac{c}{8}\partial_A (C_{BC}C^{BC})\right)+O(r^{-2}). 
\end{equation}

We consider asymptotically nonradiative vacuum (Ricci-flat) spacetimes with a Weyl tensor that approaches zero when $u \rightarrow \pm \infty$ (we denote these limits by $\mathscr I^+_\pm$). 
The boundary metric and shear in these limits are~\cite{Compere:2018ylh}
\begin{equation} \label{eq:C_AB}
\gamma_{AB}=\bar \gamma_{AB}(x^C), \qquad C_{AB}=(u+\frac{C^\pm}{c})N_{AB}^{\text{vac}}-2D_A D_B C^\pm+\gamma_{AB}D^C D_C C^\pm+O(u^{-1})  \, ,
\end{equation}
respectively.
The ``vacuum news'' $N_{AB}^{\text{vac}}(x^C)$ \cite{1977asst.conf....1G,Compere:2018ylh} is zero for round 2-sphere metrics, but it should be included when generalized BMS transformations act on the phase space.
The field $C^\pm(\theta,\phi)$ specifies the (super-)translation frame at $\mathscr I^+_\pm$, and the supertranslation transition $(D^2+2)D^2(C^+-C^-)$ is related to the displacement memory effect \cite{Zeldovich:1974aa,*Christodoulou:1991cr,*Strominger:2014pwa}. 
At $\mathscr I^+_-$ (referred to as ``spatial infinity''), the values of $N_{AB}^\mathrm{vac}$ and $C^-$ (including all harmonics) relate to a spontaneous breaking of the super-Lorentz frame (but not the Lorentz subgroup) and the (super-)translation frame, respectively; a specific Lorentz group can be associated with this fixed frame. 

The one-parameter family of definitions of super-Lorentz charges $\mathcal J^{(\alpha)}_R $ is given by~\cite{Compere:2019gft,Compere:2020lrt,Elhashash:2021iev}
\begin{equation}
\mathcal J^{(\alpha)}_R =\frac{1}{2}\int_S d\Omega R^A\left( N_A-u\partial_A \left( m + \frac{1}{8}C^{AB}N_{AB}^{\text{vac}} \right) - \frac{\alpha c^3}{4 G}C_{AB}D_C C^{BC} - \frac{\alpha c^3}{16G}\partial_A (C_{BC}C^{BC})\right) \, .\label{Ja}
\end{equation}
Angular momenta $\mathcal J^{(\alpha)}_i$ are associated with rotations $R_i^A \equiv -\epsilon^{AD}\partial_D n_i$, CM charges $\mathcal K_i^{(\alpha)}$ with boosts $K_i^A \equiv \gamma^{AD}\partial_D n_i$, and super-Lorentz charges with general smooth $R^A(x^B)$. 
Here, $\epsilon^{AB}$ is the antisymmetric tensor and $d\Omega=1/(4\pi)\sqrt{\text{det}\gamma_{AB}}d^2x$ is the area element on the celestial sphere.

For the angular momentum $\mathcal J^{(\alpha)}_i$, it was shown in~\cite{Compere:2019gft} that $\alpha=1$ corresponds to the definition proposed by Dray and Streubel~\cite{Dray:1984rfa} (its flux on $\mathscr I^+$ agrees with fluxes given by Thorne~\cite{Thorne:1980ru} and Ashtekar and Streubel~\cite{Ashtekar:1981bq}), as well as Wald and Zoupas~\cite{Wald:1999wa} (and used subsequently 
in~\cite{Barnich:2010eb,Barnich:2011mi,Chen:2013lza,*Chen:2014uma,Flanagan:2015pxa,Distler:2018rwu,Keller:2018gyt,Compere:2018ylh,Ashtekar:2019rpv,Chen:2021szm}). 
When $\alpha=0$, the definition matches that of~\cite{Pasterski:2015tva, Hawking:2016sgy,Compere:2020lrt}, and for $\alpha=3$, it matches with~\cite{1975ctf..book.....L,Bonga:2018gzr,*Damour:2020tta}.

The $\alpha$-dependent ambiguity in the angular momentum and CM charge is a distinguishable effect~\cite{Elhashash:2021iev} in the following sense: 
The flux of angular momentum and CM for generic sources will differ for different values of $\alpha$, and hence the instantaneous value of the charges will disagree even in the same fixed Bondi frame. For binary-black-hole mergers, the effect could be resolved in numerical simulations, but it is small. We now consider theoretical arguments to fix this ambiguity.

We first consider the generalized BMS charge algebra at spatial infinity given in~\cite{Compere:2020lrt}.
There the supermomentum in an arbitrary generalized BMS frame was defined as 
\begin{equation}
\mathcal P_T = \frac{1}{c}\int d\Omega \, T \; \bar m,\qquad \bar m \equiv m + \frac{1}{8}C^{AB}N_{AB}^{\text{vac}}. \label{PT}   
\end{equation}
The generalized BMS algebra of 2-sphere diffeomorphisms and supertranslations was shown in~\cite{Compere:2020lrt} to be faithfully realized by the Poisson bracket for just $\alpha =0$ in \eqref{Ja}. 
The generalized BMS charge algebra at spatial infinity is 
\begin{equation}
\lbrace \mathcal P_{T},\mathcal P_{T'} \rbrace = 0, \quad \lbrace \mathcal J^{(0)}_{R},\mathcal P_{T}\rbrace = \mathcal P_{D_{R}(T)}, \quad \lbrace \mathcal J^{(0)}_{R},\mathcal J^{(0)}_{R'} \rbrace = \mathcal J^{(0)}_{[R,R']} \, ,
\label{BMS}
\end{equation} 
where $D_{R}(T)\equiv (R^A\partial_A - \frac{1}{2} D_A R^A)T$ and where the fluxes of the generalized BMS charges match with those of~\cite{Campiglia:2020qvc}. 
Any other value of $\alpha$ [or another definition for the supermomentum \eqref{PT}] produces an anomaly in this algebra.

The generalized BMS charges with $\alpha=0$ and the supermomentum in~\eqref{PT} vanish in Minkowski spacetime even when computed in an arbitrarily supertranslated or super-Lorentz-transformed frame. 
This differs from other prescriptions~\cite{Compere:2016jwb,*Compere:2016hzt}, which have an anomalous Poisson bracket for the generalized BMS charges.
These facts both point to $\alpha=0$ being preferred.

Suppose we now quantize the charges of the generalized BMS group.
Semi-classical (canonical) quantization rules suggest to replace the classical brackets with quantum commutators over $i\hbar$ and replace the classical charges with quantum operators. 
A necessary starting point of quantization is having a faithful representation of the algebra of all the generalized BMS charges at spatial infinity.
Since the generalized BMS algebra~\eqref{BMS} is only represented for $\alpha=0$, this is another reason for favoring $\alpha=0$, which we will use henceforth. 

Unitary representations of the generalized BMS group are labeled by the eigenvalues of its Casimir invariants.
For the Lorentz quotient subgroup of the generalized BMS group, the Casimir invariant is the square of the Pauli-Lubanski spin vector; to quantize angular momentum we then need an intrinsic angular momentum that commutes with the (super-)momentum.
In non-radiative regions (including the limits of timelike infinity, $\mathscr I^+_+$, or spatial infinity, $\mathscr I^+_-$), we can fix the generalized BMS frame and determine a preferred Lorentz group of this frame (the boundary conditions \eqref{eq:C_AB} at $u\rightarrow \pm\infty$ identify two such Lorentz groups at $\mathscr I^+_\pm$).
In the CM frames at $\mathscr I^+_\pm$ with $\gamma_{AB}$ being the round-sphere metric, $C^\pm$ vanishing, and $\mathcal J_i^{(0)}$ aligning with the $z$ axis, the angular momentum $\mathcal J^{(0)}_z(\pm\infty)$ reduces to the intrinsic angular momentum of the source.
For fermionic matter, both angular momenta, $\mathcal J^{(0)}_z(\pm\infty)$, are quantized in multiples of $\hbar/2$ (and in multiples of $2\hbar$ in the absence of matter). 

We would now like to define an intrinsic angular momentum for all $u$ that does not require fixing a particular frame.
No such local expression for the angular momentum is known, but expressions have been proposed involving non-local quantities on the 2-sphere in non-generic cases~\cite{Chen:2013kza,Keller:2018gyt,Javadinazhed:2018mle,Compere:2019gft,Chen:2021szm}; none of these definitions has been shown to be consistent with generalized BMS symmetry. 
While we will not give a definitive intrinsic angular momentum here (or prove its uniqueness), we will propose an expression that is consistent with generalized BMS symmetry and behaves reasonably at $\mathscr I^+_\pm$.

Our definition comes from considering the form of the mass and angular momentum aspects for $n$ spinning, massive bodies in the limit where the distance between each particle goes to infinity as $u\rightarrow \pm\infty$.
In a super-Lorentz frame such that $N_{AB}^{\text{vac}}=0$, the mass and angular momentum aspects are given by~\cite{Compere:2019gft} 
\begin{eqnarray}
m &=& \sum_i \frac{M_{(i)}}{\gamma_{(i)}^3 \left(1- \dfrac{\vec{v_{(i)}}}{c} \cdot \vec{n}\right)^3},\qquad \gamma_{(i)}(v_{(i)}) = \frac{1}{\sqrt{1-\dfrac{v_{(i)}^2}{c^2}}} \, , 
\label{mfinal}\\
N_A &=&  \dfrac{2 m}{c}\partial_A C + \partial_A \left[m\left(u+\frac{C}{c}\right) \right]  -\sum_i \frac{3 J_{(i)} }{c^2 \gamma_{(i)}^2 \left(1-\dfrac{\vec{v_{(i)}}}{c} \cdot \vec{n}\right)^2 } \sin^2 \theta_{(i)} \partial_A \phi_{(i)} \, ,
\label{NAfinal}
\end{eqnarray}
where the index $(i)$ labels the individual bodies and $M_{(i)}$, $J_{(i)}$ and $v_{(i)}$ are their respective masses, angular momenta and velocities ($\theta_{(i)}$ and $\phi_{(i)}$ specify the boost directions).
This expression suggests that the intrinsic (super-)Lorentz charge \cite{Compere:2019gft} should take the following form at $\mathscr I^+_\pm$:
\begin{equation} \label{eq:JRinv}
    \mathcal J_R^{\text{inv}}\equiv \mathcal J_R^{(0)} - \mathcal P_{D_{R}C^\pm}=\frac{1}{2}\int_S d\Omega R^A\left( N_A-u\partial_A\bar m -\frac{2 \bar m}{c} \partial_A C^\pm -\frac{1}{c}\partial_A (\bar m C^\pm) \right) \, . 
\end{equation}
One can easily check at $\mathscr I^+_\pm$ that
\begin{eqnarray}
\{\mathcal J_R^{\text{inv}} , \mathcal P_T\} = 0,\qquad \{ \mathcal J_{R}^{\text{inv}} ,  \mathcal J_{R'}^{\text{inv}} \}  = \mathcal J_{[R,R']}^{\text{inv}} \, .
\label{alginv}
\end{eqnarray}
Thus, $\mathcal J_R^{\text{inv}}$ is manifestly supertranslation invariant and (super-)Lorentz covariant at $\mathscr I^+_\pm$. 
Because $C^-$ is obtained by inverting an elliptic operator on the 2-sphere from gradients of local 2-sphere tensors, it is a non-local quantity, as anticipated in \cite{Javadinazhed:2018mle}. 
In a fixed CM frame, the angular momentum is similar to that of~\cite{Chen:2021szm} at $\mathscr I^+_\pm$, but we fixed $\alpha=0$ (\cite{Chen:2021szm} uses $\alpha=1$) to accommodate generalized BMS symmetries.  

Inspired by~\cite{Chen:2021szm}, the simplest expression for the Lorentz charges at finite $u$ which is consistent with generalized BMS symmetry is given by 
\begin{equation}
\mathcal J_R^{\text{int}}(u) =\frac{1}{2}\int_S d\Omega R^A\left( N_A-u\partial_A\bar m -\frac{2 \bar m}{c} \partial_A C-\frac{1}{c}\partial_A (\bar m C) \right) \, , \label{challenge}
\end{equation}
where $C(u,x^A)$ is now defined from the decomposition $C_{AB}=(u+\frac{C}{c})N_{AB}^{\text{vac}}-2D_A D_B C+\gamma_{AB}D^2C+\epsilon_{C(A}D^C D_{B)} \psi$, $N_{AB}^{\text{vac}}$ is determined from $\bar\gamma_{AB}(x^C)$ and $\bar m$ is given in~\eqref{PT}.

Finally, to resolve the CM ambiguity, we must identify the $SO(3)$ subgroup of the (super-)Lorentz group on cuts $u$ that are the rotation generators in the CM frame.
They can be inferred from either \eqref{NAfinal} and \eqref{challenge}, or~\cite{Ashtekar:2019rpv} to be 
\begin{equation}
R^{\prime A}_i = \gamma R_i^A +(1-\gamma)\frac{v_i (\vec{v}\cdot \vec{R}^A)}{v^2}+\gamma \epsilon_{ijk}v_j K_k^A, \qquad v_i \equiv \frac{\mathcal P_i}{\mathcal P_0},\qquad \gamma=(1-\frac{v^2}{c^2})^{-\frac{1}{2}},\quad v=\sqrt{\vec{v}\cdot \vec{v}} \,. \label{Ri}
\end{equation}
Since the momentum $\mathcal P_\mu$ commutes with $\mathcal J_R^{\text{inv}}$, the commutation rules \eqref{alginv} still hold for $R=R^{\prime A}_i$ at $\mathscr I^+_\pm$. The angular momentum~\eqref{challenge} for the vector~\eqref{Ri} is our proposal for the quantized intrinsic angular momentum. 

The realization of the algebra of generalized BMS charges in~\eqref{BMS} at spatial infinity suggested that the $\alpha$-dependent ambiguity in definitions could be resolved, thereby straightening out this latest twist in the story of angular momentum in general relativity.
Aiming to quantize the angular momentum initiated a new turn in this narrative by suggesting that non-local terms will be needed to obtain an intrinsic angular momentum that admits quantization.
We anticipate that new spins on angular momentum will appear in the future.

\vspace{5pt} \noindent {\bf Acknowledgments.} G.C. would like to thank the organizers and participants of the ``Flat Asymptotia workshop'' for interesting discussions. G.C. is a Research Associate of the F.R.S.-FNRS. G. C. acknowledges support from the FNRS research credit No. J003620F, the IISN convention No. 4.4503.15, and the COST Action GWverse No. CA16104. D.A.N.\ acknowledges support from the NSF grant PHY-2011784.\vspace{-10pt}


%

\end{document}